\documentclass[conference]{IEEEtran}
\IEEEoverridecommandlockouts

\usepackage{cite}
\usepackage{amsmath,amssymb,amsfonts}
\usepackage{algorithmic}
\usepackage{graphicx}
\usepackage{textcomp}
\usepackage{xcolor}
\def\BibTeX{{\rm B\kern-.05em{\sc i\kern-.025em b}\kern-.08em
    T\kern-.1667em\lower.7ex\hbox{E}\kern-.125emX}}
\begin{document}

\title{Indoor Localization Based on MSC Map\\

}
\author{\IEEEauthorblockN{1\textsuperscript{st} \L ukasz Ku\l acz}
\IEEEauthorblockA{\textit{Poznan University of Technology}\\
Poznan, Poland \\
lukasz.kulacz@put.poznan.pl}
\and
\IEEEauthorblockN{2\textsuperscript{nd} Adrian Kliks}
\IEEEauthorblockA{\textit{Poznan University of Technology}\\
Poznan, Poland}
\and
\IEEEauthorblockN{3\textsuperscript{nd} Julius Ruseckas}
\IEEEauthorblockA{\textit{Baltic Institute of Advanced Technology} \\
Vilnius, Lithuania}
\and
\IEEEauthorblockN{4\textsuperscript{nd} Gediminas Molis}
\IEEEauthorblockA{\textit{Baltic Institute of Advanced Technology} \\
Vilnius, Lithuania}
}

\maketitle

\begin{abstract}
In this short paper, we propose a technique for AI-based identification of modulation and coding schemes (MCS) in surrounding cellular signals. Based on the created MCS map, we evaluate the performance of indoor localization techniques.  
\end{abstract}

\begin{IEEEkeywords}
AI, MCS, indoor localization
\end{IEEEkeywords}

\section{Introduction}
\label{sec:intro}
Indoor localization belongs to one of the important research domains in the context of contemporary wireless networks and signal processing. Typically, the solutions based on Global Navigation Satellite Systems (GNSS) work very well outdoors, yet they do not offer enough accuracy and reliability for indoor applications. Thus, various methods for indoor positioning have been proposed including fingerprinting, pre-stored maps of certain parameters, or received signal strength from many sources,  \cite{Guo2019,Jang2019}. 
In this paper, we consider the application of supervised learning to detect the location of indoor users. Intentionally, we assume that no GNSS signals can be used and there is access only to the legacy signals (like 4G or 5G) from the nearby base stations. In particular, we assume that the detection of the so-called modulation and coding schemes (MCS) could be used to specify the user's location. We propose to train the neural network  (NN) in such a way that it will be able to detect the MCS values in the legacy 4G/5G signals. Once detected, the pre-stored map of MCSes can be used to specify the true location of the user by comparing the measured/detected MCS with those stored in the map. Such an approach allows us to decide on the location of the users just by observing the signal and without decoding it. Here we assume that no additional sources of information are used; however, achieved results are the basis for future extensions, where we foresee the usage of Reconfigurable Intelligent Surfaces (RISes) to improve signal detection and localization. 

\section{Detection of Modulation and Coding Scheme}
In the experiments that were conducted, we assume the usage of NN for MCS detection. In practice, the value of MCS is selected dynamically based on the observed and measured quality of the transmission channel. As the transmission environment is typically time-varying, and the users (UE) are often moving, the values of MCS may vary frequently. According to the 5G standards, it can take a value between 0 to 31, and these numbers are translated to various signal modulation and coding setups, thus - to different achievable throughputs. Based on the channel observation between the UE and base stations (BS), the proper value of MCS is selected. Thus, in our supervised approach, we have to label the signal for various MCS values. 

For that purse, we have built the experimentation setup consisting of the legacy 4G/5G BS provided by the Amarisoft company and off-the-shelf smartphones. To train the NN, in the repeatable scenario (i.e., for the known and fixed location of the UE and the base station receiver), we have gathered I-Q samples of the received signal from the UE at the BS for different MCS values. The Amarisoft BS allows us to enforce (by adjusting the software) the fixed value of MCS for a certain transmission. In our experiment, the uplink direction was considered. In reality, the received signal strength at the BS will vary, mainly depending on the distance between the UE and BS. To properly train the NN, the I-Q samples for each considered MCS have been collected for various signal-to-interference-plus-noise ratios (SINRs), with the approximate step of 1 dB. The samples were gathered by the FSL6 spectrum analyzer by Rohde \& Schwarz, whose antenna was collocated with that of the smartphone.  The sampling frequency has been set to 5 MHz, and at each signal run 523776 samples have been detected and saved in a file with a single precision format. Ten files were created for each combination of MCS and SINR. The Amarisoft BS operated at the 2.68 GHz central frequency, with the 1.4 MHz channel bandwidth (equivalent to six resource blocks). In order to reflect the real signal (i.e., the data frames filled with user data), at the UE side, the dedicated commercial application for testing the maximum throughput of the connection was set. Such collected data have been post-processed in Python while training the specified network configuration. Finally, we used the signals with MCS from 8 to 16. As mentioned, for each MCS-SINR tuple, there are ten files, which have been split into one signal for the validation dataset and the remaining nine signals for the training dataset. From each signal, we randomly select 1000 samples with a length of 2048 for experiments.

In detail, for signal classification, we use a model similar to the temporal convolution network (TCN). The neural network uses 1D fully-convolutional architecture. However, since for classification, there is no need to ensure that the future does not influence the past, we do not use causal convolutions. Each hidden layer is the same length as the input layer; zero padding is added to keep subsequent layers the same length as previous ones. The neural network uses dilated convolutions to achieve an exponentially large receptive field. The dilation factor $d$ is increased exponentially with the depth of the network: $d=2i$ at level $i$ of the network. As in the TCN model, we use residual blocks where the output of a block is formed as a sum of the block input and non-linear transformations of the input. Such a residual block allows layers to learn modifications to the identity mapping rather than the entire transformation, leading to easier training of very deep networks. Within a residual block, the neural network has two layers of dilated convolution and rectified linear unit (ReLU) non-linearities. We also use an additional 1x1 convolution to transform the input of the residual block before element-wise addition. For the classification task, we take the output of the last residual block and perform global average pooling, the output of which is transformed by the densely connected hidden layer with ReLU nonlinearity and the output densely connected layer with softmax activation. Categorical cross-entropy is used as a loss function of the network. We use the neural network with kernel size $k=5$, number of filters in the hidden layers $N=64$, and $n=8$ residual blocks (leading to the dilation factor of the last block $d=28=256$). Such a neural network has 355854 trainable parameters.
We train the model using the AdamW optimizer with the weight decay $10^{-2}$. The learning rate was changed using a 1cycle learning rate schedule with the maximum learning rate of $10^{-2}$. The neural network was trained for 50 epochs.

The true experiment has been prepared to test the detection ability of the trained network. The same setup has been used (i.e., the Amarisoft BS, off-the-shelf smartphone, and the collocated spectrum analyzer R\&S FSL6) to collect the I-Q samples of the real signal transmitted between the UE and BS, where the selection of MCS has been made dynamically, following the standard procedures. However, at the same time, we were able to record the true values of the chosen MCS based on analysis of the logs. The trained network has been used to detect the MCS values of the recorded signal. The achieved accuracy was at the level of 81\%.


\section{Experiment results for indoor localization }
In the main experiment, we used the analogous setup as previously; mainly, the same Amaisoft legacy base station has been used inside the room at the premises of Poznan University of Technology. The whole room has been split into 54 tiles ($6\times 9$) of 1 m size, and for each tile, the averaged MCS value for typical transmission has been measured. These measurements have been used to create the map of the average MCS inside the room, as shown in Fig. 2, where axes indicate the tile indexes and colors of the MCS value. 
\begin{figure}[ht]
  \centering
  \includegraphics[width=0.75\linewidth]{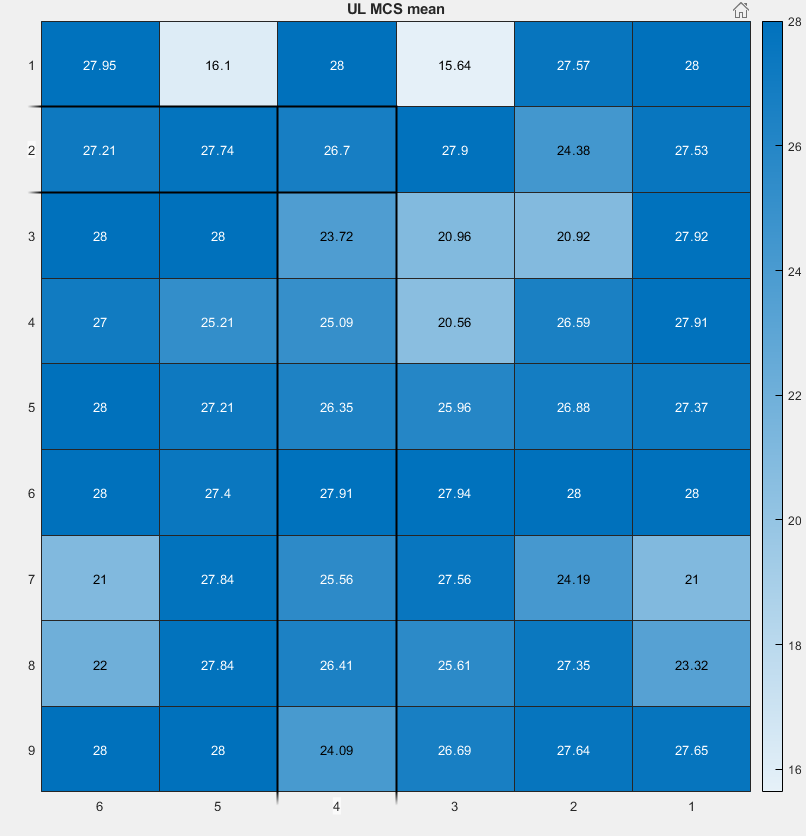}
  \caption{MCS map covering the area of one laboratory}
\end{figure}
In the main experiment, the same trained NN was used to detect the MCS and, finally, after postprocessing - to indicate the location inside the room. The achieved results are shown in the form of the confusion matrix in Fig. 3. One can see that the results are fine, yet to guarantee precise localization, more processing will be needed in the future, for example, by applying advanced dead-reckoning techniques or to consider the usage of RISes.  

\begin{figure}[ht]
  \centering
  \includegraphics[width=0.8\linewidth]{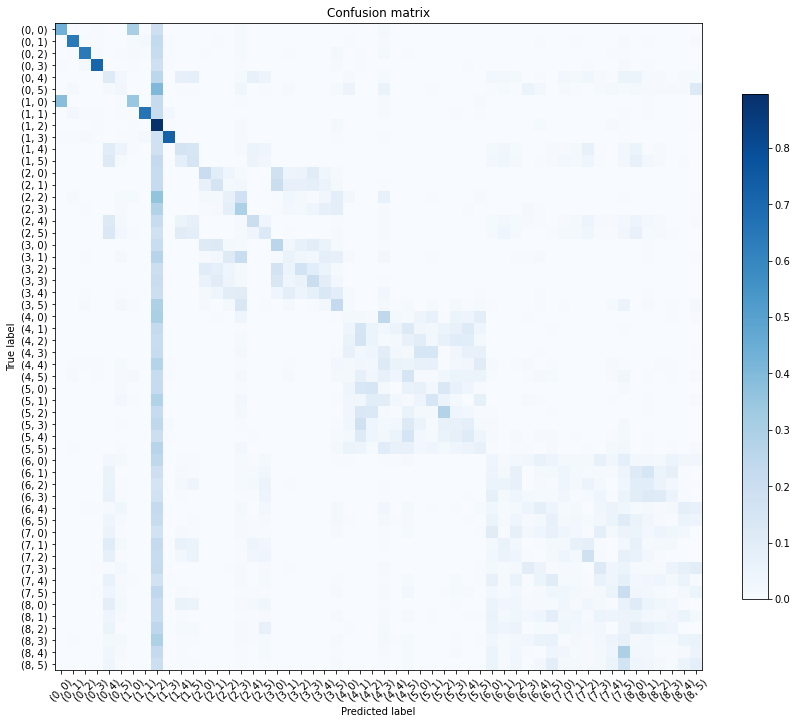}
  \caption{Confusion matrix for indoor localization}
\end{figure}


\section*{Acknowledgment}
The work has been realized within $2017/27/\text{L}/\text{ST7}/03166$ project, and the continued work by L. Kukacz and A.~Kliks has been realized within project no. $2021/43/\text{B}/\text{ST}7/01365$ funded by National Science Center in~Poland.
\bibliographystyle{IEEEtran}
 \bibliography{sample-base}

\begin{thebibliography}{1}
\providecommand{\url}[1]{#1}
\csname url@samestyle\endcsname
\providecommand{\newblock}{\relax}
\providecommand{\bibinfo}[2]{#2}
\providecommand{\BIBentrySTDinterwordspacing}{\spaceskip=0pt\relax}
\providecommand{\BIBentryALTinterwordstretchfactor}{4}
\providecommand{\BIBentryALTinterwordspacing}{\spaceskip=\fontdimen2\font plus
\BIBentryALTinterwordstretchfactor\fontdimen3\font minus
  \fontdimen4\font\relax}
\providecommand{\BIBforeignlanguage}[2]{{%
\expandafter\ifx\csname l@#1\endcsname\relax
\typeout{** WARNING: IEEEtran.bst: No hyphenation pattern has been}%
\typeout{** loaded for the language `#1'. Using the pattern for}%
\typeout{** the default language instead.}%
\else
\language=\csname l@#1\endcsname
\fi
#2}}
\providecommand{\BIBdecl}{\relax}
\BIBdecl

\bibitem{Guo2019}
G.~Guo, R.~Chen, F.~Ye, X.~Peng, Z.~Liu, and Y.~Pan, ``Indoor smartphone
  localization: A hybrid wifi rtt-rss ranging approach,'' \emph{IEEE Access},
  vol.~7, pp. 176\,767--176\,781, 2019.

\bibitem{Jang2019}
B.~Jang and H.~Kim, ``Indoor positioning technologies without offline
  fingerprinting map: A survey,'' \emph{IEEE Com. Surveys \& Tutorials},
  vol.~21, no.~1, pp. 508--525, 2019.

\end{thebibliography}

\end{document}